\documentclass[12pt]{article}
\usepackage{amssymb,amsmath,epsfig, float}
\allowdisplaybreaks
\begin{document}

\title{\bf Study of Decoupled Gravastars in Energy-momentum Squared Gravity}
\author{M. Sharif$^1$ \thanks {msharif.math@pu.edu.pk} and Saba Naz$^2$
\thanks{sabanaz1.math@gmail.com}\\
$^1$ Department of Mathematics and Statistics, The University of Lahore,\\
1-KM Defence Road Lahore, Pakistan.\\
$^2$ Department of Mathematics, University of the Punjab,\\
Quaid-e-Azam Campus, Lahore-54590, Pakistan.}

\date{}

\maketitle
\begin{abstract}
In this paper, we generate an exact anisotropic gravastar model
using gravitational decoupling technique through minimal geometric
deformation in the framework of $f(\Re, {T}^{2})$ gravity. This
novel model explains an ultra-compact stellar configuration whose
internal region is smoothly matched to the exterior region. The
developed stellar model satisfies some of the essential
characteristics of a physically acceptable model such as a positive
monotonically decreasing profile of energy density from the center
to the boundary and monotonically decreasing behavior of the
pressure. The anisotropic factor and Schwarzschild spacetime follows
physically acceptable behavior. We find that all the energy bounds
are satisfied except strong energy condition inside the
ultra-compact stellar structure for the coupling constant of this
theory, which is compatible with the regularity condition.
\end{abstract}
\textbf{Keywords:} Modified theories; Decoupling; Gravastars;
Israel formalism.\\
\textbf{PACS:} 04.50.Kd; 04.40.-b; 97.10.Cv.

\section{Introduction}

The current accelerated expansion of the cosmos has been the most
significant development in recent years. This expansion is thought
to be the result of a mysterious force having repulsive nature and
is described as dark energy (DE). Several researchers have put
forward their efforts to reveal its unknown features. The first
proposal explaining the hidden characteristics of DE is the
cosmological constant, but it suffers issues like fine-tuning and
coincidence. Different modified versions have been presented to
overcome these problems that are assumed to be fascinating
approaches in revealing mysteries of the universe. The first
modification is $f(\Re)$ theory of gravity which has a significant
literature \cite{1} to understand its physical properties. This
modified theory was further generalized by incorporating the idea of
curvature-matter coupling. Such coupling theories are non-conserved
with an additional force that alters the path of the particle. The
minimal coupling is introduced as $f(\Re, T)$ theory \cite{2}
whereas the non-minimally coupled theory is $f(\Re, T,
\Re_{\sigma\gamma}T^{\sigma\gamma})$ gravity \cite{3}.

Different cosmic studies have been presented to understand the
beginning of the universe. One of the extensively accepted proposals
termed as big-bang theory is a remarkable framework describing
evolutionary processes. According to this idea, all the matter in
the cosmos expanded from a single point, referred to as a
singularity. This theory portrays the beginning of the universe but
this proposal suffers flatness problem, horizon problem and monopole
problem. The research community is always curious and dedicates its
effort to find answers to cosmological issues. The fascinating
notion of bounce theory (based on repeated cosmic expanding and
contracting behavior) serves as a backbone for introducing a new
theory.  This newly introduced concept resolves big bang issues by
resolving singularity and provides a better explanation of
accelerated expanding cosmic behavior.

Katrici and Kavuk \cite{35} constructed an extension of $f(\Re)$
theory that incorporates the concept of bounce theory. They
developed a particular coupling between matter and gravity through
self-contraction of the energy-momentum tensor (EMT) referred to as
energy-momentum squared gravity (EMSG) or known as $f(\Re, {T}^{2})$
theory, where $T^{2}=T_{\sigma\gamma}T^{\sigma\gamma}$. This
proposal with a minimum scale factor as well as finite maximum
energy density is considered to be a suitable framework to overcome
big-bang problem. The cosmological constant resolves the big-bang
singularity by providing the repulsive force in the background of
this theory. This theory follows the true sequence of cosmological
eras and effectively describes cosmic behavior. The field equations
involve squared and product components of matter variables, which
are useful in studying different cosmological scenarios.

Roshan and Shojai \cite{36} computed the exact solution by solving
EMSG field equations with homogeneous isotropic spacetime and
demonstrated the possibility of a bounce in the early universe.
Board and Barrow \cite{37} calculated the range of exact solutions
for an isotropic expanding universe in reference to early and
late-time evolution. Nari and Roshan \cite{38} determined a
connection between mass and the radius of neutron stars. They found
that a smaller or larger value of mass is governed by the central
pressure of these stars and the value of model parameter of this
gravity. Bahamonde et al. \cite{41} used different coupling models
to study the expansion of the universe and concluded that these
models help in understanding the present accelerated cosmic
expansion. We have studied various attributes of charged as well as
uncharged gravastar solutions \cite{44}. Recently, Sharif and his
collaborators studied decoupled solutions \cite{44*} as well as the
impact of charge on complexity of static sphere in the same theory
\cite{44**}.

Different cosmic phenomena, such as the origin and evolution of
celestial bodies have captivated the interest of various
researchers. Among all the cosmic objects, stars are the core
components of galaxies, which are organized systematically in a
cosmic web. When a star runs out of fuel, its outward pressure
vanishes, leading to gravitational collapse and hence compact
objects are formed. A black hole is such a stellar remnant which is
a totally collapsed entity with a singularity hidden behind an event
horizon. Mazur and Motolla \cite{13m} developed a compact model
(gravastar) as an alternative to black hole to avoid singularity and
event horizon. Motolla \cite{ppp} discussed in detail dark energy
and condensate stars. The resulting gravitational condensate star
configuration resolves all black hole paradoxes, and provides a
testable alternative to black holes as the final state of complete
gravitational collapse. Mazur and Motolla \cite{ttt} studied
pressure and negative pressure interior of a non-singular black
hole. In contrast to black holes, the significant feature of this
hypothetical object is its singularity-free nature. Three regions
constitute the complete gravastar structure, internal and external
regions separated by a thin shell. The DE in the inner domain causes
a repulsive force which contributes as a main barrier to overcome
singularity formation. The intermediate shell surrounding the inner
boundary exerts an inward force and thus hydrostatic equilibrium is
maintained while the Schwarzschild metric characterizes the exterior
region. Moreover, each region is described by a specific equation of
state (EoS).

Visser and Wiltshire \cite{14m} examined the stability of gravastars
towards radial perturbations and concluded that a viable EoS results
in the stability of gravastar in GR. This work was extended by
examining appropriate constraints for the stability of gravastar
solutions \cite{15m}. Cattoen et al. \cite{15mm} studied the usual
gravastar structures and concluded that the cosmic configuration has
anisotropy in the absence of intermediate shell. Bilic et al.
\cite{15mmm} formulated gravastar solutions by replacing the inner
Born-Infeld phantom metric with the de Sitter geometry which
represent compact objects at the galactic center. Horvat and Ilijic
\cite{16m} discussed the stability of gravastars by employing the
speed of sound criteria on thin-shell to determine compactness
bounds. Ovalle \cite{n} discussed anisotropic gravastars by means of
decoupling approach.

Many researchers have also investigated the formation as well as
fundamental physical characteristics of gravastars in modified
theories of gravity. Das et al. \cite{29aa} discussed isotropic
gravastar structure in the realm of ${f(\Re, T)}$ gravity. They
obtained a linear profile of physical features in relation to shell
thickness. Shamir and Ahmad \cite{29tt} constructed spherically
symmetric gravastar models whose characteristics obey an increasing
trend with respect to the thickness. Different physical features of
charged/uncharged gravastar model were studied in $f(\mathbb{T})$
theory ($f(\mathbb{T})$ is an arbitrary function of the torsion
scalar) \cite{29aaa}. Abbas and Majeed \cite{65aaa} discussed
isotropic gravastar structure in the background of Rastall gravity.
Yousaf et al. \cite{65aa} examined the gravastar model in modified
theory and found that length, energy and entropy present an
increasing trend with respect to thickness. In the background of
${f(\Re, \mathcal{G})}$ gravity ($\mathcal{G}$ defines the
Gauss-Bonnet invariant), Bhatti et al. \cite{28} investigated
different physical characteristics related to intrinsic shell
thickness and found accepted behavior. Ray et al. \cite{rr} provided
a very good review on the entire aspects of gravastar as envisioned
by Mazur and Mottola starting from the black hole physics to its
present state and enlightening the future works. The gravastar
models in a number of modified gravity models starting from $f(R,T)$
to Rastall-Rainbow gravity have been reviewed. Bhatti and his
collaborators \cite{28a} discussed charged/uncharged gravastar model
in ${f(\mathcal{G})}$ gravity to analyze different features. Bhar
and Rej \cite{cg77} studied the role of electromagnetic field in the
stability of gravastar structure.

Cosmological solutions are very important to comprehend the
structural properties as well as the mechanism of celestial
structures. However, it is usually difficult to obtain exact
solutions due to the presence of highly non-linear terms in the
field equations. In this regard, the gravitational decoupling
technique by minimal geometric deformation (MGD) has recently been
developed to determine the astrophysical and cosmic solutions.
Ovalle \cite{11} firstly introduced this concept from the viewpoint
of braneworld. This approach connects a new gravitational source to
the EMT of the seed matter distribution through a non-dimensional
parameter. In this scheme, the deformation is employed only on the
radial metric function while leaving the temporal metric component
unaltered. Consequently, two sets of nonlinear field equations are
formed, one for the new source and the other for the seed source.
The decoupling method has extensively been used to transform
isotropic spherical solutions into anisotropic ones.

The isotropic, anisotropic, and charged fluid configurations are the
key topics for researchers in the context of compact celestial
bodies. Ovalle et al. \cite{12} extended isotropic solutions by
assuming Tolman IV spacetime in the inner region. Ovalle and his
collaborators \cite{12*} investigated ultra-compact stars through
decoupling technique without altering any of the fundamental
property. Gabbanelli et al. \cite{13} computed acceptable
anisotropic solutions by considering the Durgapal-Fuloria stellar
configuration in the interior region. Graterol \cite{14} implemented
this technique to determine anisotropic solutions by considering
Buchdahl isotropic solution. The extensions of isotropic solutions
were obtained by using Krori-Barua (KB) spacetime \cite {15}.
Morales and Tello-Ortiz \cite{17} studied charged Heintzmann
anisotropic solution and graphically examined matter variables for
various stars. Maurya et al. \cite{18} explored the influence of
anisotropy by taking Korkina-Orlyanskii isotropic matter
distribution.

Many viable and stable solutions have been studied by using the
decoupling technique in modified theories. Sharif and Majid
\cite{46} obtained anisotropic solutions by using different
well-known isotropic solutions in Brans-Dicke theory. Maurya et al.
\cite{301} explored anisotropic solutions by assuming
Korkina-Orlyanskii spacetime in the inner domain in $f(\Re, T)$
theory. Zubair and Azmat \cite{303} extended Tolman VII solution in
the same theory. Sharif and Naseer \cite{306} evaluated
charged/uncharged anisotropic extensions of KB metric in non-minimal
curvature-matter coupled theory. Recently, Azmat et al. \cite{n1}
employed this technique to investigate physical features of
anisotropic gravastar in $f(\Re, T)$ theory. The departure from the
isotropic scenario leads to an interesting situation, giving rise to
an intriguing phenomenon inside the stellar interior, i.e., the
system experiences a repulsive force that counteracts the
gravitational gradient, which allows the construction of more
compact and massive objects. This inspired us to develop anisotropic
gravastars in the background of $f(\Re, T^2)$ providing interesting
results.

The aim of this article is to study the anisotropic gravastar in the
background of $f(\Re, {T}^{2})$ theory. We use gravitational
decoupling through MGD scheme to include the impact of anisotropy in
the fluid configuration of the gravastar. The paper is organized as
follows. Section \textbf{2} demonstrates the complete array of the
field equations corresponding to both sources (seed and source). The
solutions are decoupled in section \textbf{3} which are smoothly
joined at the stellar surface. Section \textbf{4} describes MGD
gravastar configuration which meets some of the viable
characteristics of a stellar structure. We summarize our results in
the last section.

\section{Basic Formalism}

The action of this theory involving matter Lagrangian density
($\L_m$) and the Lagrangian density ($\L_{\tau}$) of an additional
source ($\tau$) is characterized by
\begin{equation}\label{1}
S=\int d^{4}x\left(\frac{1}{2k^{2}}f(\Re,T^{2})+ {\L}_m +\aleph
{\L}_{\tau}\right)\sqrt{-g}.
\end{equation}
The coupling constant is $k^{2}=8\pi$, determinant of the metric
tensor is denoted by $g$ and $\aleph$ symbolizes the decoupling
parameter. The action (\ref{1}) provides the following field
equations
\begin{eqnarray}\label{2}
\Re_{\sigma\gamma}f_\Re+g_{\sigma\gamma}\Box
f_\Re-\nabla_{\sigma}\nabla_{\gamma}f_\Re-\frac{1}{2}g_{\sigma\gamma}f
=8\pi\left(T_{\sigma\gamma}+\aleph\tau_{\sigma\gamma}\right)-f_{T^{2}}\circleddash_{\sigma\gamma},
\end{eqnarray}
here $f_\Re=\frac{\partial{f}}{\partial{\Re}}$ and
$f_{T^{2}}=\frac{\partial{f}}{\partial{T^{2}}}$. The self
contraction of del operator defines d'Alembert operator
($\Box=\nabla_{\sigma}\nabla^{\sigma}$), whereas the tensor
$\circleddash_{\sigma\gamma}$ is
\begin{equation}\label{3}
\circleddash_{\sigma\gamma}=-2{\L}_m(T_{\sigma\gamma}
-\frac{1}{2}g_{\sigma\gamma}T)
-4\frac{\partial^{2}{{\L}_m}}{\partial{g^{\sigma\gamma}}{\partial{g^{\alpha\delta}}}}
T^{\alpha\delta}-TT_{\sigma\gamma}+2T^{\delta}_{\sigma}T_{\gamma\delta}.
\end{equation}
The EMT of seed and additional sources are, respectively
\begin{equation}\label{4}
T_{\sigma\gamma}=g_{\sigma\gamma}{\L}_{m}-2\frac{\partial
{\L}_{m}}{\partial g^{\sigma\gamma}},\quad
\tau_{\sigma\gamma}=g_{\sigma\gamma}{\L}_{\tau}-2\frac{\partial
{\L}_{\tau}}{\partial g^{\sigma\gamma}}.
\end{equation}
The isotropic matter configuration involving density, pressure and
four-velocity corresponds to the seed source given as follows
\begin{equation}\label{6}
T_{\sigma\gamma}=(\varrho+P)U_{\sigma}U_{\gamma}-g_{\sigma\gamma}P,
\end{equation}
with $U_{\gamma}U^{\gamma}=1$. Moreover, plugging $L_{m}=-P$ in
Eq.(\ref{3}) yields
\begin{equation}\label{6*}
\circleddash_{\sigma\gamma}=(3P^{2}+\varrho^{2}+4P\varrho)U_{\sigma}U_{\gamma}.
\end{equation}
Rearrangement of Eq.(\ref{2}) gives
\begin{equation}\label{8}
G_{\sigma\gamma}=\frac{1}{f_\Re} \Bigg( 8\pi(T_{\sigma\gamma}
+\aleph \tau_{\sigma\gamma})-g_{\sigma\gamma}\Box
f_{\Re}+\nabla_{\sigma}\nabla_{\gamma}f_{\Re}-
\circleddash_{\sigma\gamma}f_{T^{2}}+\frac{1}{2}g_{\sigma\gamma}(f-\Re
f_{\Re})\Bigg),
\end{equation}
where $G_{\sigma\gamma}=\Re_{\sigma\gamma}-\frac{1}{2}\Re
g_{\sigma\gamma}$ is the Einstein tensor.

The static spherical object is described by the following line
element
\begin{equation}\label{9}
ds^{2}=e^{\varphi}dt^{2}-e^{\psi}dr^{2}-r^{2}d\theta^{2}-r^{2}\sin^{2}\theta
d\phi^{2},
\end{equation}
where the metric coefficients $\varphi$ and $\psi$ are functions of
radial coordinate only. The complexity of the field equations is
reduced by considering minimally coupled model $f(\Re,
T^{2})=\Re+\zeta T^{2}$ ($\zeta$ is the coupling parameter) and the
respective equations take the following form
\begin{eqnarray}\label{10}
8\pi(\widetilde{\varrho}+\aleph\tau^{t}_{t})&=&e^{-\psi}
(\frac{\psi^{'}}{r}-\frac{1}{r^{2}})+\frac{1}{r^{2}},\\\label{11}
8\pi(\widetilde{P}-\aleph\tau^{r}_{r})&=&e^{-\psi}(\frac{\varphi^{'}}{r}
+\frac{1}{r^{2}})-\frac{1}{r^{2}},\\\label{12}
8\pi(\widetilde{P}-\aleph\tau^{\theta}_{\theta})&=&\frac{e^{-\psi}}{4}(2\varphi^{''}
-\frac{2\psi^{'}}{r}+\frac{2\varphi^{'}}{r}-
\varphi^{'}\psi^{'}+\varphi^{'2}).
\end{eqnarray}
Here prime exhibits radial derivative whereas
($\widetilde{\varrho}$, $\widetilde{P}$) are provided by
\begin{eqnarray}\label{13}
\widetilde{\varrho}&=&\varrho-\frac{1}{16\pi}(3P^{2}+8\varrho
P+\varrho^{2})\zeta,\\\label{14}
\widetilde{P}&=&P-\frac{1}{16\pi}(3P^{2}+\varrho^{2})\zeta.
\end{eqnarray}
The conservation law is violated in this theory and hence covariant
differentiation of Eq.(\ref{8}) gives
\begin{equation}\label{15}
\nabla^{\sigma}(\Theta_{\sigma\gamma}f_{T^{2}})-\frac{1}{2}
g_{\sigma\gamma}\nabla^{\sigma}f=
P^{'}+\frac{\varphi^{'}}{2}(\varrho+P)+\aleph\left(\frac{\varphi^{'}}{2}
(\tau^{t}_{t}-\tau^{r}_{r})\right.+
\left.\frac{2}{r}(\tau^{\theta}_{\theta}-\tau^{r}_{r})-\tau^{r'}_{r}\right).
\end{equation}
The highly non-linear field equations involve seven unknown
functions: state variables $(\widetilde{\rho}, \widetilde{P})$,
metric coefficients $(\varphi,\psi)$ and components of extra source
$(\tau^{t}_{t},\tau^{r}_{r},\tau^{\theta}_{\theta})$. The total
energy density, radial/tangential pressure are respectively, given
as
\begin{equation}\label{16}
\hat{\varrho}=\widetilde{\varrho}+\aleph\tau^{t}_{t},\quad
\hat{P_{r}}=\widetilde{P}-\aleph\tau^{r}_{r},\quad
\hat{P_{t}}=\widetilde{P}-\aleph\tau^{\theta}_{\theta}.
\end{equation}
Moreover, the presence of an additional source with
$\tau^{r}_{r}\neq\tau^{\theta}_{\theta}$ generates anisotropy in the
inner domain of self-gravitating body having the form
\begin{equation}\label{19}
\hat\Delta=\hat{P_{t}}-\hat{P_{r}}=\aleph(\tau^{r}_{r}-\tau^{\theta}_{\theta}).
\end{equation}
If $\zeta\rightarrow0$, the field equations in $f(\Re, T^{2})$
gravity will reduce to GR.

\section{Minimal Geometric Decoupling Scheme}

In this section, the solution of the field equations are discussed
through MGD approach. To encode the influence of extra source in
isotropic matter distribution, the metric potentials ($\xi$ and
$\varpi$) are modified in the following manner
\begin{equation}\label{20}
\xi\rightarrow\varphi=\xi+\aleph j(r), \quad \varpi\rightarrow
e^{-\psi}=\varpi+\aleph \varsigma,
\end{equation}
where $j(r)$ and $\varsigma(r)$ present temporal and radial
deformations, respectively. Employing MGD scheme (alters only radial
function through geometric deformation while keeping the temporal
metric function unchanged, i.e., $j(r)=0$), we have two sets of the
decoupled solutions. The first array is given by
\begin{eqnarray}\label{22}
8\pi\widetilde{\varrho}&=&\frac{1}{r^{2}}-(\frac{\varpi}{r^{2}}+\frac{\varpi^{'}}{r})=8\pi\varrho-
\frac{1}{2}(\varrho^{2}+3P^{2}+8\varrho P)\zeta, \\\label{23}
8\pi\widetilde{P}&=&-\frac{1}{r^{2}}+\frac{\varpi}{r}(\frac{1}{r}+\varphi^{'})=8\pi
P-\frac{1}{2} (\varrho^{2}+3P^{2})\zeta,\\\label{24}
8\pi\widetilde{P}&=&\frac{\varpi}{2}(\varphi^{''}+\frac{\varphi^{'2}}{2}+\frac{\varphi^{'}}{r})
+\frac{\varpi^{'}}{2}(\frac{\varphi^{'}}{2}+\frac{1}{r})= 8\pi
P-\frac{1}{2}(\varrho^{2}+3P^{2})\zeta,
\end{eqnarray}
whereas, the second array including the impact of extra source is
\begin{eqnarray}\label{25}
8\pi\tau^{t}_{t}&=&-\frac{\varsigma^{'}}{r}-\frac{\varsigma}{r^{2}},\\\label{26}
8\pi\tau^{r}_{r}&=&-\frac{\varsigma}{r}(\frac{1}{r}+\varphi^{'}),\\\label{27}
8\pi\tau^{\theta}_{\theta}&=&-(\frac{\varphi^{'}}{r}+\varphi^{''}+
\frac{\varphi^{'2}}{2})\frac{2\varsigma}{r}-\varsigma^{'}
(\frac{\varphi^{'}}{4}+\frac{1}{2r}).
\end{eqnarray}
Equation (\ref{15}) can be splitted into isotropic and anisotropic
sectors as
\begin{eqnarray}\label{28}
\nabla^{\sigma}(\circleddash_{\sigma\gamma}f_{T^{2}})
-\frac{1}{2}g_{\sigma\gamma}\nabla^{\sigma}f&=&(\widetilde{P})^{'}+
\frac{\varphi^{'}}{2}(\widetilde{\varrho}+\widetilde{P}),
\\\label{29}
\tau^{r'}_{r}-\frac{\varphi^{'}}{2}(\tau^{t}_{t}-\tau^{r}_{r})
-\frac{2}{r}(\tau^{\theta}_{\theta}-\tau^{r}_{r})&=&0.
\end{eqnarray}
We consider a standard barotropic EoS to simplify the field
equations as
\begin{equation}\label{30}
\varrho=\frac{m}{V},
\end{equation}
where $m$ and $V=\frac{4}{3}\pi{r}^{3}$ describe the mass and volume
of the sphere. Using this EoS, the matter variables (density and
pressure) read as follows
\begin{eqnarray}\label{31}
\widetilde\varrho&=&\frac{3 m \left(-9 \zeta  m+16 \pi ^2
r^3\right)}{64 \pi ^3 r^6}, \\\label{32} \widetilde{P}&=&\frac{48
\pi ^2 m r^3-9 \zeta  m^2}{64 \pi ^3 r^6},
\end{eqnarray}
which contribute considerably in studying the influence of
anisotropy.

\subsection{Matching Condition}

The deformed Schwarzschild spacetime describing the exterior region,
is modified trough $\tau$-sector and is described as \cite{12*}
\begin{equation}\label{36}
ds^{2}_{+}=\left(1-\frac{2\mathcal{M}}{r}\right)dt^{2}-\left(1-\frac{2\mathcal{M}}{r}
+\mathcal{V}g^{\star}(r)\right)^{-1}dr^{2}
-r^{2}d\theta^{2}-r^{2}\sin^{2}\theta d\phi^{2},
\end{equation}
The decoupling function for the exterior Schwarzschild metric is
$g^{\star}(r)$, which appears to the extra source
${\tau_{\sigma\gamma}^{+}}$. It is mentioned here that the internal
$\varsigma$ and external $g^{\star}(r)$ deformations are not
identical. Likewise, the respective parameters $\aleph$ and
$\mathcal{V}$ are not identical and govern the deformations in their
respective sectors. The anisotropic source
${\tau_{\sigma\gamma}^{+}}\neq0$ may impact exterior geometry and it
is no more vacuum. Moreover, when decoupling is employed to the
Schwarzschild vacuum as well as additional external source,
Eqs.(\ref{25})-(\ref{27}) for the external geometry give
\begin{eqnarray}\label{36*}
(\tau^{t}_{t})^{+}&=&-\frac{g^{\star}}{r^{2}}-\frac{(g^{\star})^{'}}{r},\\\label{36**}
(\tau^{r}_{r})^{+}&=&-\frac{g^{\star}}{r(r-2\mathcal{M})},\\\label{36***}
(\tau^{\theta}_{\theta})^{+}&=&\frac{\mathcal{M}(r-\mathcal{M})}{r^{2}(r-2\mathcal{M})}g^{\star}
-\frac{r-\mathcal{M}}{2r(r-2\mathcal{M})}(g^{\star})^{'}.
\end{eqnarray}
Junctions conditions are essential for joining the inner and outer
geometries at the hypersurface $(\Sigma:r=\mathfrak{R})$. Israel
junction conditions have two fundamental forms. The first form reads
the continuity of metric potentials
\begin{equation}\label{3600}
({ds^{2}})_\Sigma=0.
\end{equation}
At stellar surface, we have
\begin{equation}\label{360}
e^\varphi=1-\frac{2\mathcal{M}}{\mathfrak{R}},
\end{equation}
and
\begin{equation}\label{361}
1-\frac{2M}{\mathfrak{R}}+\aleph\varsigma(r)=1-\frac{2\mathcal{M}}{\mathfrak{R}}+\mathcal{V}g^{\star}(r).
\end{equation}
while the second form is
\begin{equation}\label{36**}
(T^{1}_{1})^{-1}-\aleph(\tau^{r}_{r})^{-}=-\mathcal{V}(\tau^{r}_{r})^{+}.
\end{equation}
This condition leads to the continuity of pressure in the radial
direction at the hypersurface.

\section{MGD Gravastars in EMSG}

Following Mazur and Mottola gravastar model \cite{13m}, we use the
EoS, $\mathcal{W}\varrho=P$ with $\mathcal{W}$ describing the EoS
parameter. The inner gravastar domain follows DE EoS $\varrho=-P$
(for $\mathcal{W}=-1$). The negative pressure in this region
provides repulsive force. Employing DE EoS along with Eq.(\ref{15})
yields $\varrho=\varrho_{0}$(constant) which gives
\begin{equation}\label{21}
\varpi=1-\frac{2{\rho_{0}}{r^{2}}}{3}(4\pi+\zeta{\varrho_{0}})+\frac{C_{1}}{r},
\end{equation}
In order to avoid singularity, we take the arbitrary constant
$C_{1}$ to be zero at core. Thus we have
\begin{equation}\label{22}
\varpi=1-\frac{2{\varrho_{0}}{r^{2}}}{3}(4\pi+\zeta{\varrho_{0}}).
\end{equation}
The temporal and radial metric functions are connected as follows
\begin{equation}\label{23}
e^{-\varphi}={C_{2}}\varpi,
\end{equation}
The mass corresponding to the inner region is
\begin{equation}\label{24}
M(\mathfrak{R})=\frac{{\varrho_{0}(4\pi+\zeta{\varrho_{0}})
}\mathfrak{R}^{3}}{3}.
\end{equation}
Applying the deformation metric potentials in explicit form, we have
\begin{eqnarray}\label{23*}
e^{\varphi}&=&{C_{2}}(1-{\mathcal{P}}^{2}r^{2}),\\
\varpi&=&1-{\mathcal{P}}^{2}r^{2},
\end{eqnarray}
where
\begin{equation}\label{24*}
\mathcal{P}^{2}=\frac{\varrho_{0}}{3}\left(8\pi+2\zeta\varrho_{0}\right)=
\frac{2\mathcal{M}}{\mathfrak{R}^{3}}=\frac{\mathfrak{R}_{s}}{\mathfrak{R}^{3}}.
\end{equation}
Here, $\mathcal{M}$ is the mass and $\mathfrak{R}$ denotes radius of
the star. Moreover, $\mathfrak{R}_s$ is the Schwarzschild radius and
$\varrho_0$ is the constant energy density.

The deformed gravastar solution is developed by using the MGD
approach yielding
\begin{eqnarray}\label{101*}
e^{\varphi}&=&{C_{2}}(1-{\mathcal{P}}^{2}r^{2}),\\\label{bb*}
e^{-\psi}&=&1-{\mathcal{P}}^{2}r^{2}+\aleph \varsigma(r).
\end{eqnarray}
We solve the system of equations (\ref{25})-(\ref{27}) to obtain the
deformation function. To determine the remaining components of extra
source, the characteristic of gravastar ($g_{00} = g_{11}^{-1}=0$)
at the stellar surface $r = \mathfrak{R}_s$ is assumed. To preserve
this property in the deformed metric components of gravastar
((\ref{101*}) and (\ref{bb*})), the geometric deformation function
satisfies
\begin{equation}\label{102}
\varsigma(r)\sim1-{\mathcal{P}}^{2}r^{2}.
\end{equation}
The simplest choice is
\begin{eqnarray}\label{103}
\varsigma(r)\sim(1-{\mathcal{P}}^{2}r^{2}){\mathcal{P}}^{n}r^{n},~~~~n\geq2.
\end{eqnarray}
This condition on $n$ is very significant to obtain non-singular
solutions. The expression for the radial metric component
(\ref{bb*}) is evaluated as
\begin{equation}\label{104}
e^{-\psi}=(1-{\mathcal{P}}^{2}r^{2})({1+\aleph\mathcal{P}}^{n}r^{n}).
\end{equation}
We take a constraint, $\aleph\geq-1$, to have finite positive value
at $r=\mathfrak{R}_s$. Consequently, we obtain matter variables as
follows
\begin{eqnarray}\label{106}
\hat{\varrho}&=&\widetilde{\varrho}+\aleph\mathcal{P}^{n}r^{n-2}\left((n+3){\mathcal{P}}^{2}r^2-n-1\right),\\
\hat{P_{r}}&=&\widetilde{P}-\aleph{\mathcal{P}}^{n}{r}^{n-2}\left(3{\mathcal{P}}^{2}{r}^2-1\right),\\\label{107}
\hat{P_{t}}&=&\widetilde{P}-\aleph{\mathcal{P}}^{n}{r}^{n-2}\Big[\left(n+3\right){\mathcal{P}}^{2}{r}^2-\frac{n}{2}\Big].\label{108}
\end{eqnarray}
It can be easily seen from Eqs.(\ref{106})-(\ref{108}) that only
$n\geq2$ provides singularity-free solutions, whereas the
anisotropic factor is given by
\begin{equation}\label{19}
\hat\Delta=\hat{P_{t}}-\hat{P_{r}}=\frac{\aleph{\mathcal{P}}^{n}{r}^{n-2}}{2}\left(n-2-2n\mathcal{P}^{2}r^2\right).
\end{equation}

The expressions (\ref{104})-(\ref{108}) along with (\ref{101*})
represent an ultra-compact non-uniform anisotropic gravastar
structure. Further, we consider gravastar solution obtained after
employing deformation, given in Eqs.(\ref{101*}) and (\ref{104})
along with Eqs.(\ref{360}) and (\ref{361}) providing continuity of
metric potentials implying that
\begin{equation}\label{107}
1-\frac{2\mathcal{M}}{\mathfrak{R}}=0,
\end{equation}
\begin{equation}\label{109}
1-\frac{2\mathcal{M}}{\mathfrak{R}}+\mathcal{V}g^{\star}_{\mathfrak{R}}=0.
\end{equation}
The primary feature of the gravastar configuration
($g_{00}=g_{11}^{-1}=0$ at Schwarzschild radius) gives zero in the
above equations. Equation (\ref{109}) yields
\begin{equation}\label{111}
g^{\star}_{\mathfrak{R}}\sim\frac{2\mathcal{M}-\mathfrak{R}}{\mathfrak{R}}=0
\end{equation}
Equation (\ref{111}) shows that the deformation disappears at the
surface of gravastar geometry
\begin{equation}\label{1111}
\frac{48 \pi ^2 M r^3-27 \zeta  M^2}{64 \pi ^3 r^6}-\frac{2\aleph}
{r^2}=\frac{g^{\star}_{\mathfrak{R}}}{\mathfrak{R}-2\mathcal{M}}.
\end{equation}
Equations (\ref{107}), (\ref{111}) and (\ref{1111}) give the
necessary and sufficient constraints for smooth matching of internal
(\ref{9}) and external geometries (\ref{36}). Furthermore,
considering the standard Schwarzschild spacetime, we obtain the
decoupling parameter
\begin{equation}\label{112}
\aleph=\frac{3\left(16\pi^2Mr^3-9\zeta M^2\right)}{128\pi^3r^4}.
\end{equation}

Figure \textbf{1} shows acceptable behavior of state variables for
this value of $\aleph$. We see that density has positive decreasing
profile while the pressure is negative. The inner gravastar geometry
possesses negative pressure providing repulsive behavior. The
anisotropic factor follows decreasing trend. To obtain the exterior
deformed solution, we consider
\begin{figure}\center
\epsfig{file=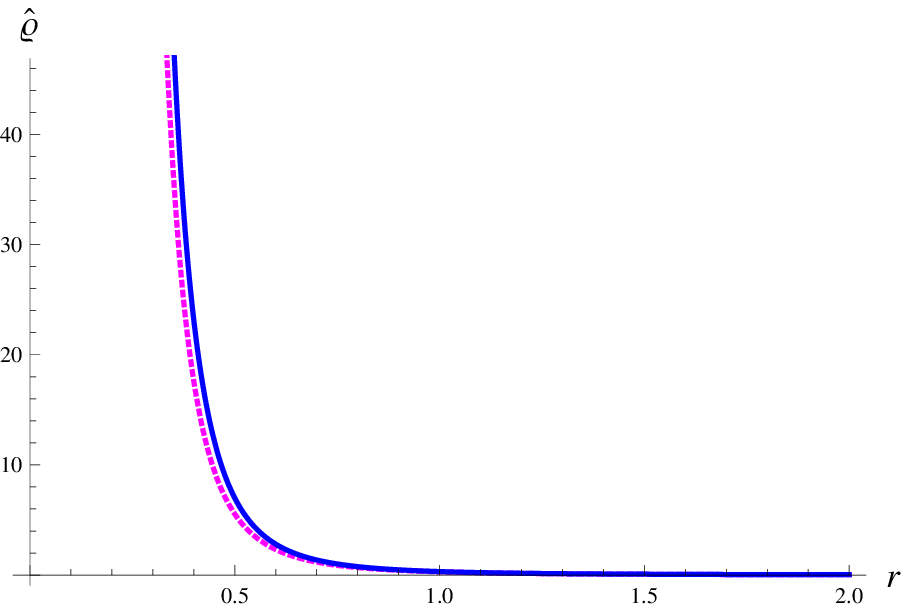,width=0.45\linewidth}\epsfig{file=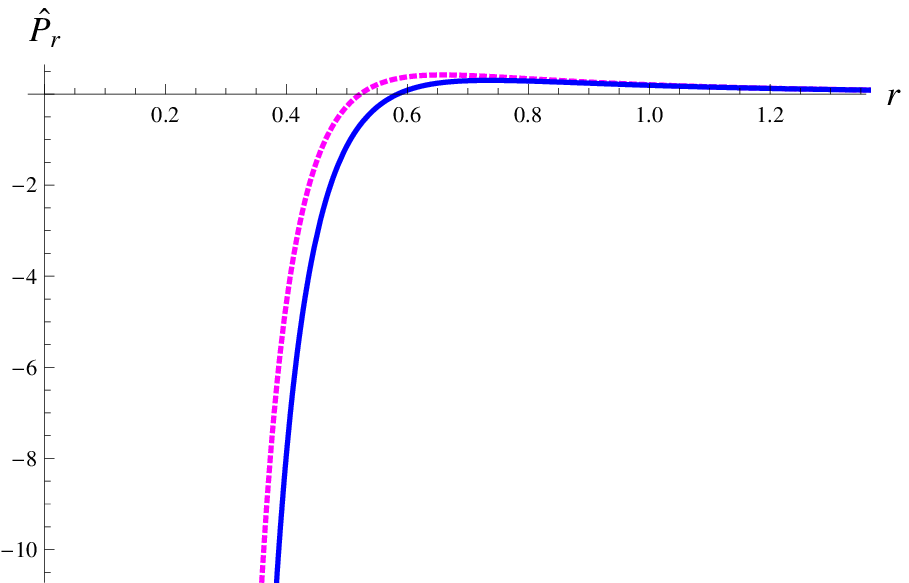,width=0.45\linewidth}
\epsfig{file=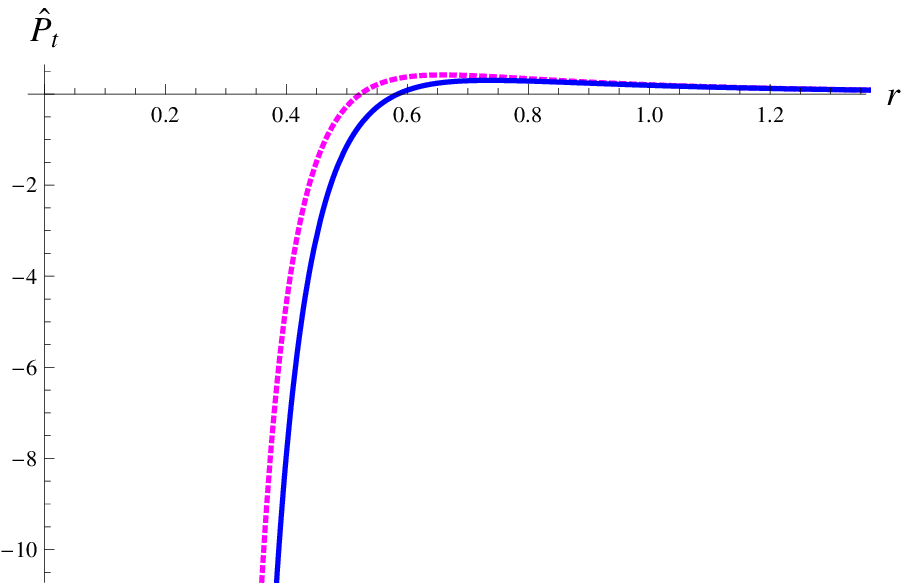,width=0.45\linewidth}\epsfig{file=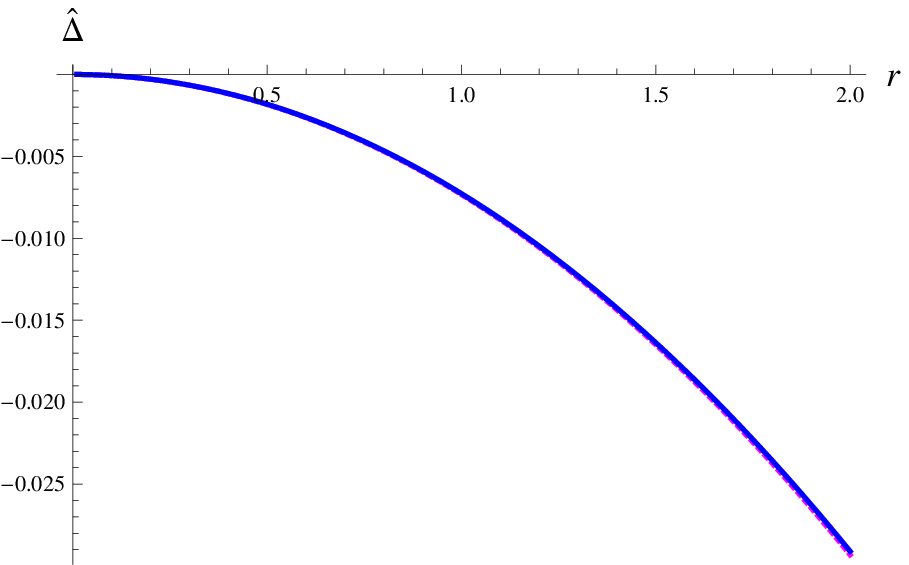,width=0.45\linewidth}
\caption{Plots of density, radial/tangential pressure and anisotropy
for $\aleph=2.5$ (Magenta) and $\aleph=3.5$ (Blue) of the interior
region.}
\end{figure}
\begin{equation}\label{113}
{\tau^{\sigma}_{\sigma}}^{+}=0.
\end{equation}
Using Eqs.(\ref{36*})-(\ref{36***}) in the above relation, we obtain
\begin{equation}\label{113}
r(6 \mathcal{M}^{2}+2r^{2}-7\mathcal{M}r){\varsigma}'(r)+2(3
\mathcal{M}^{2}+r^{2}-4\mathcal{M}r)\varsigma(r)=0,
\end{equation}
whose solution yields
\begin{equation}\label{113*}
g^{\star}(r)=\frac{l_c \left(1-\frac{2 M}{r}\right)}{2 r-3 M}.
\end{equation}
Here, $l_c$ denotes constant having dimension of length. Thus, the
deformed exterior solution is given as
\begin{equation}\label{114}
e^{-\psi}=\left(1-\frac{2\mathcal{M}}{\mathfrak{R}}\right)\left(1+\frac{l}{2r-3\mathcal{M}}\right).
\end{equation}
For $r>>\mathcal{M}$, it takes the form
\begin{equation}\label{115}
e^{-\psi}\sim1+\frac{l-4\mathcal{M}}{2r}.
\end{equation}
The matter variables corresponding to the exterior region become
\begin{eqnarray}\label{117}
(\hat{\varrho})^{+}=(\tau^{t}_{t})^{+}&=&-\frac{l\mathcal{M}}{{\left(2r-3\mathcal{M}\right)^{2}}r^2},\\
(\hat{P_{r}})^{+}=(\tau^{r}_{r})^{+}&=&\frac{l}{{\left(2r-3\mathcal{M}\right)r^2}},\\
(\hat{P_{t}})^{+}=(\tau^{\theta}_{\theta})^{+}&=&-\frac{l\left(r-\mathcal{M}\right)}
{{\left(2r-3\mathcal{M}\right)^{2}}r^2}.
\end{eqnarray}
The exterior metric anisotropy turns out to be
\begin{figure}\center
\epsfig{file=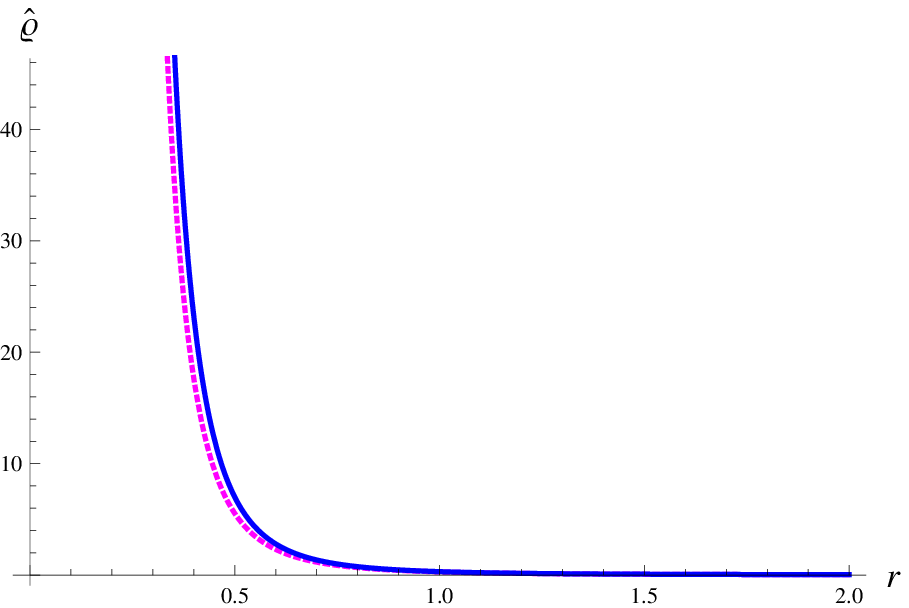,width=0.45\linewidth}\epsfig{file=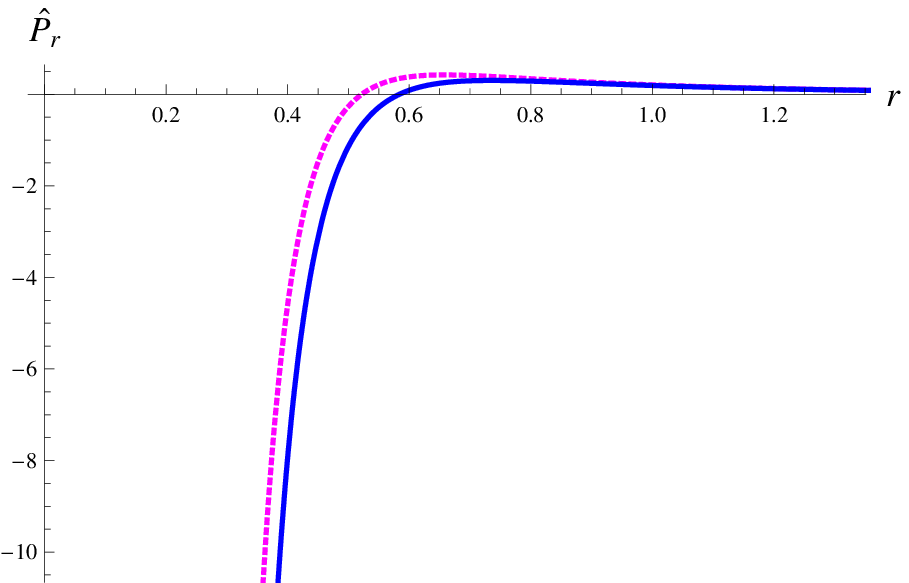,width=0.45\linewidth}
\epsfig{file=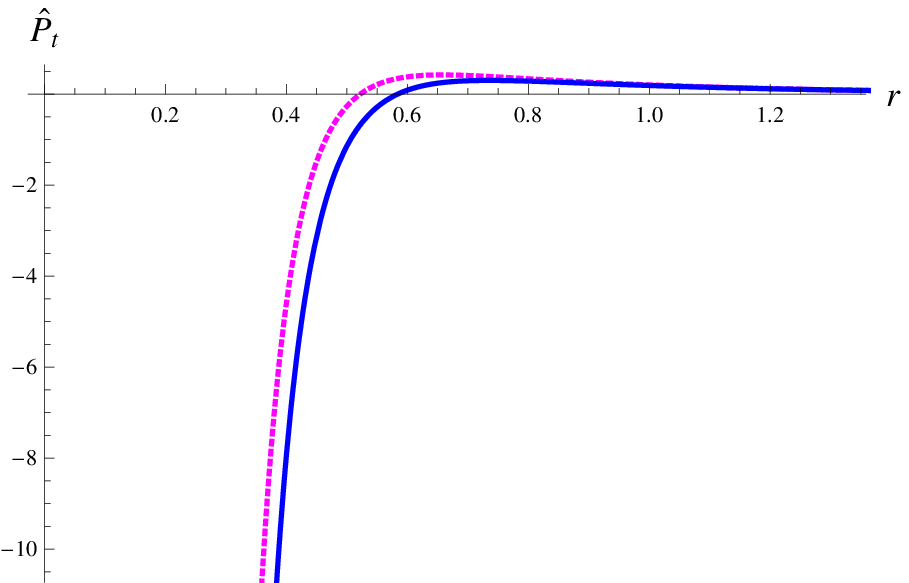,width=0.45\linewidth}\epsfig{file=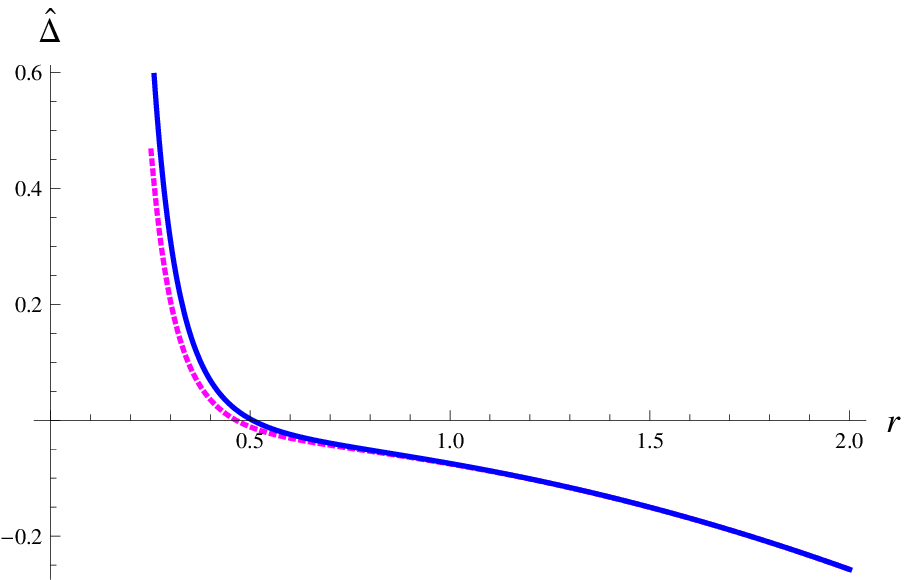,width=0.45\linewidth}
\caption{Behavior of density, radial/tangential pressure and
anisotropy for $\aleph=2.5$ (Magenta) and $\aleph=3.5$ (Blue) of the
exterior region.}
\end{figure}
\begin{equation}\label{116}
(\hat\Delta)^{+}=\frac{l(3r-4\mathcal{M})}{\left(2r-3\mathcal{M}\right)^{2}r^2}.
\end{equation}
The deformed external metric will be regular for the following value
of decoupling parameter
\begin{equation}\label{116}
\aleph=\frac{-27 \zeta  M^2+48 \pi ^2 M r^3+64 \pi ^3 r^6}{128 \pi^3
r^6},\quad\\l=-\mathcal{M}
\end{equation}
The restriction $n\geq2$ is very significant as it allows smooth
matching of the internal region with the external. The behavior of
state variables corresponding to the above decoupling parameter and
anisotropy is plotted in Figure \textbf{2}. We find monotonically
decreasing profile of density, negative pressure, and positive
anisotropy.

The left and right plots in Figure \textbf{3} indicate Schwarzschild
and deformed Schwarzschild metrics, respectively. The behavior of
deformed Schwarzschild spacetime is physically acceptable but is
totally different from that of standard Schwarzschild (linear
profile). Figure \textbf{4} provides positive finite behavior of the
metric functions of the inner and outer regions and cusp like trend
is observed at joining point. This shows non-analytic behavior which
implies violation of the second junction condition. From the general
formalism of the Israel junction conditions, a violation of the
second junction condition implies the presence of a
$\delta$-distribution of stresses located on the hypersurface
$\mathfrak{R}_{s}$. Finally, we investigate some physical
characteristics of anisotropic solutions. The existence of ordinary
matter as well as viability of the obtained solutions can be ensured
by applying energy bounds which are classified into weak, dominant,
null as well as strong energy conditions. For anisotropic matter
distribution, these energy bounds are described by
\begin{figure}\center
\epsfig{file=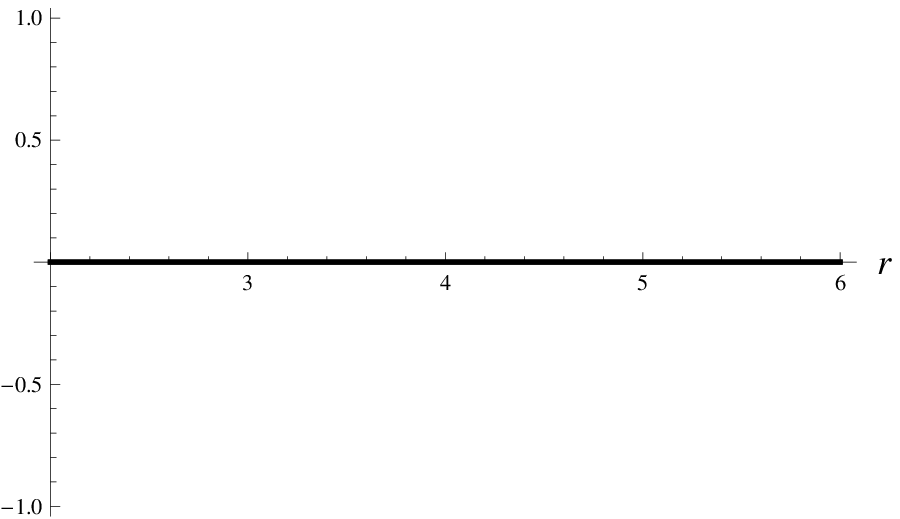,width=0.45\linewidth}
\epsfig{file=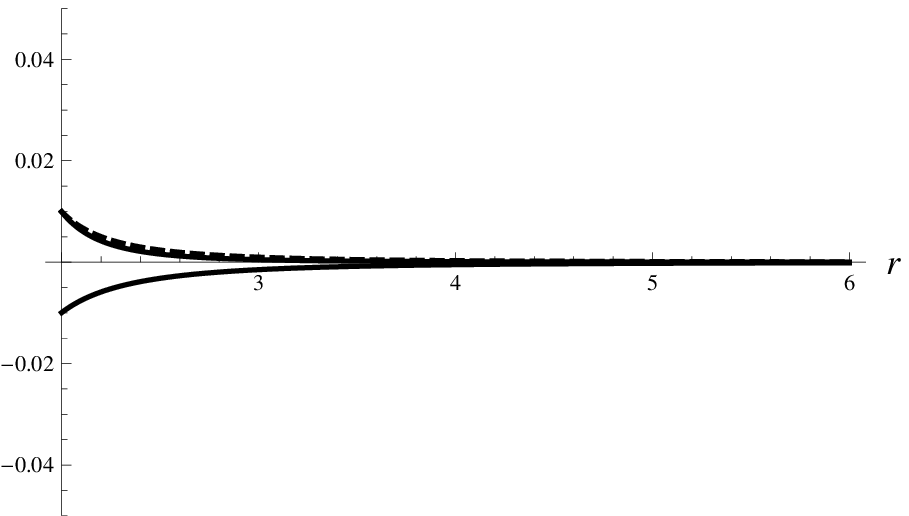,width=0.35\linewidth} \caption{Plots of the
standard Schwarzschild (left) and deformed Schwarzschild (right)
exterior.}
\end{figure}
\begin{figure}\center
\epsfig{file=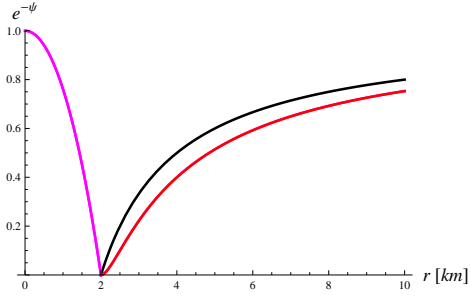,width=0.45\linewidth} \caption{Graphical trend
of the metric functions - interior function (Magenta), standard
Schwarzschild (Black) and deformed Schwarzschild (Red).}
\end{figure}
\begin{eqnarray}\nonumber
&&\text{WEC:}\quad\hat{\varrho}\geq0,\quad
\hat{\varrho}+\hat{P_{r}}\geq0,\quad \hat{\varrho}+\hat{P_{t}}\geq0,
\\\nonumber
&&\text{DEC:}\quad\hat{\varrho}-|\hat{P_{r}}|\geq0,\quad
\hat{\varrho}-|\hat{P_{t}}|\geq0,\\\nonumber
&&\text{NEC:}\quad\hat{\varrho}+\hat{P_{r}}\geq0,\quad
\hat{\varrho}+\hat{P_{t}}\geq0,
\\\nonumber
&&\text{SEC:}\quad\hat{\varrho}+\hat{P_{r}}\geq0,\quad
\hat{\varrho}+\hat{P_{t}}\geq0,\quad
\hat{\varrho}+\hat{P_{r}}+2\hat{P_{t}}\geq0.
\end{eqnarray}
Figures \textbf{5} and \textbf{6} represent that all energy bounds
except SEC are satisfied and thus the resulting solution follows
physical viability.
\begin{figure}\center
\epsfig{file=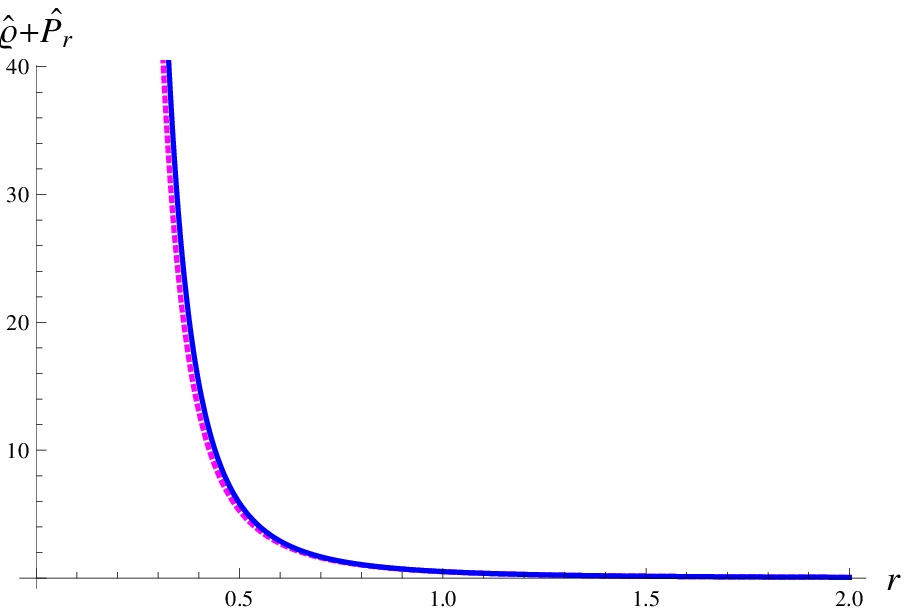,width=0.45\linewidth}\epsfig{file=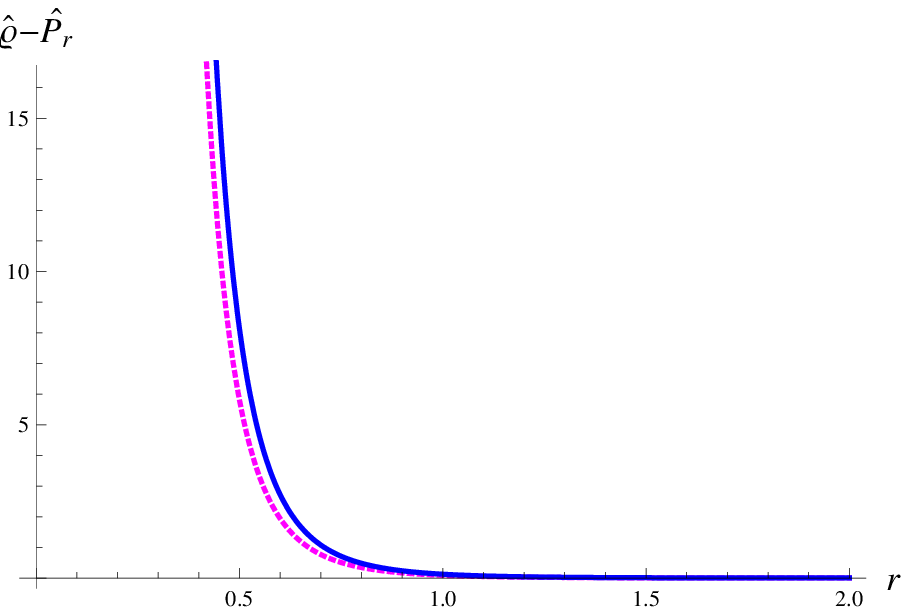,width=0.45\linewidth}
\epsfig{file=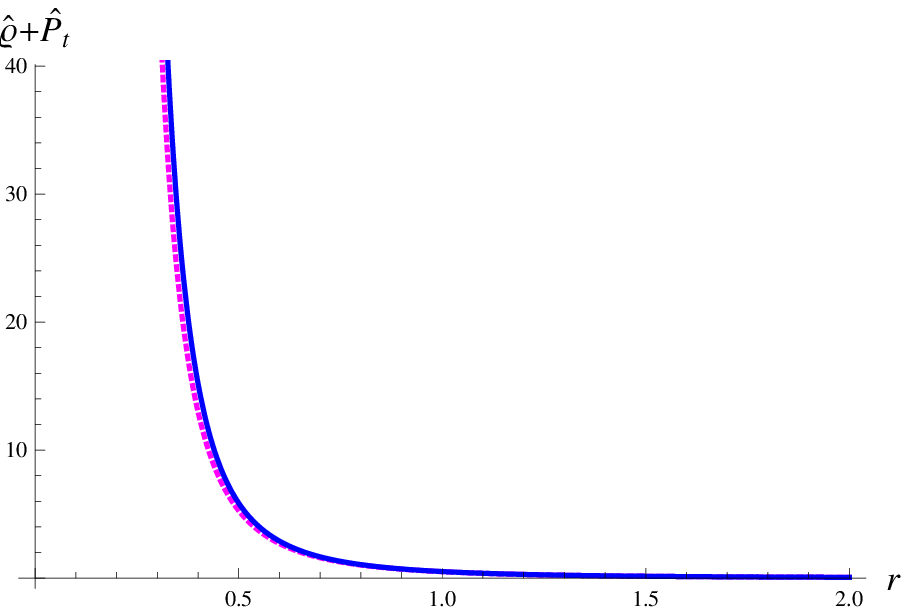,width=0.45\linewidth}\epsfig{file=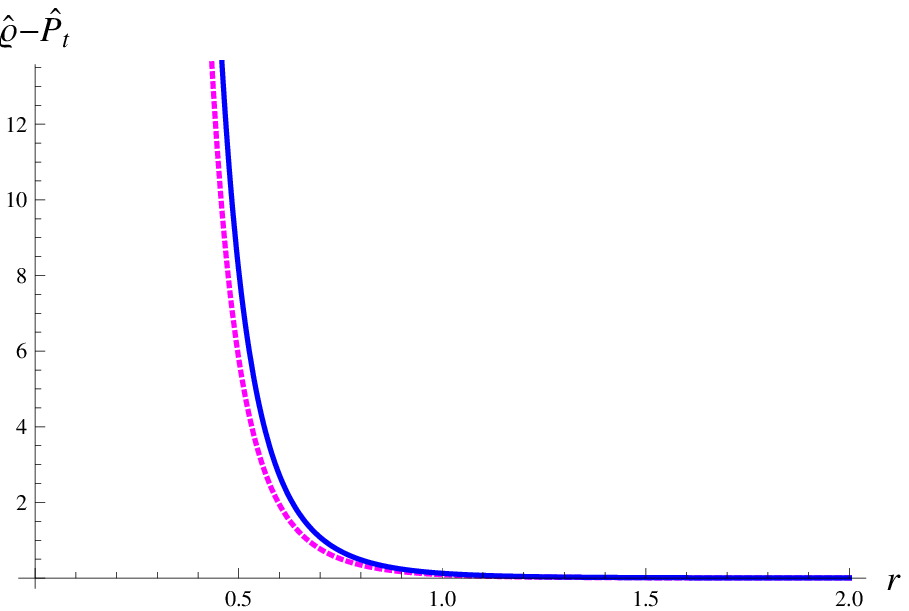,width=0.45\linewidth}
\epsfig{file=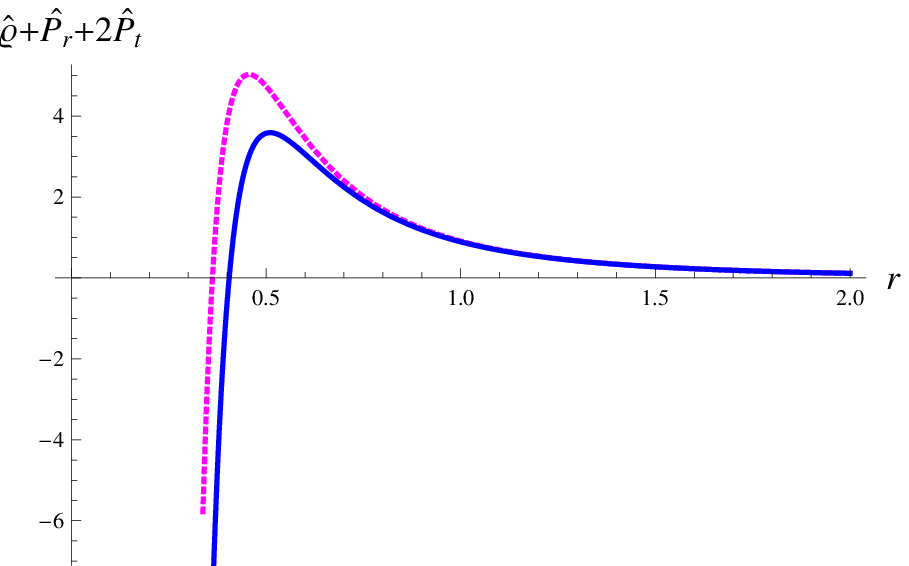,width=0.45\linewidth} \caption{Energy bounds of
the interior for $\aleph=2.5$ (Magenta) and $\aleph=3.5$ (Blue).}
\end{figure}
\begin{figure}\center
\epsfig{file=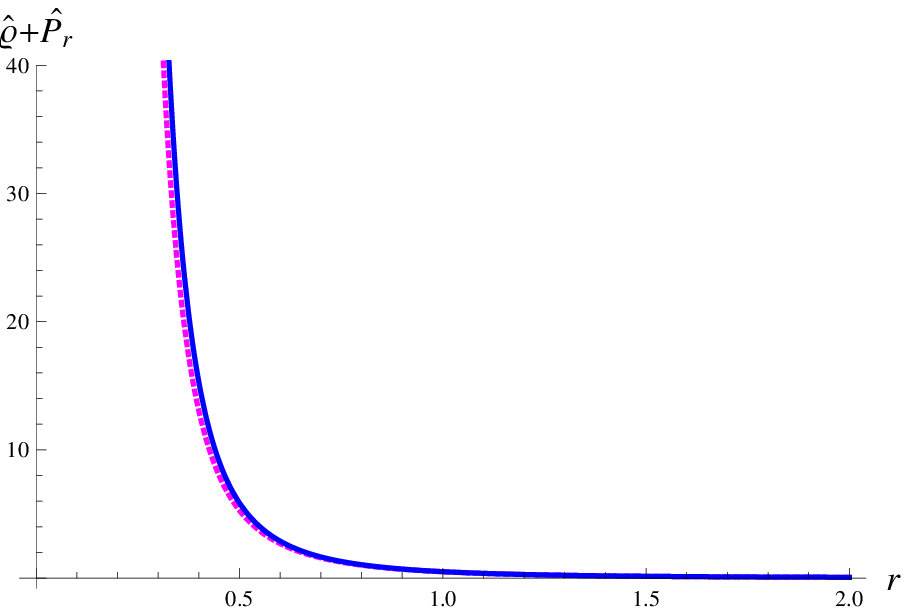,width=0.45\linewidth}\epsfig{file=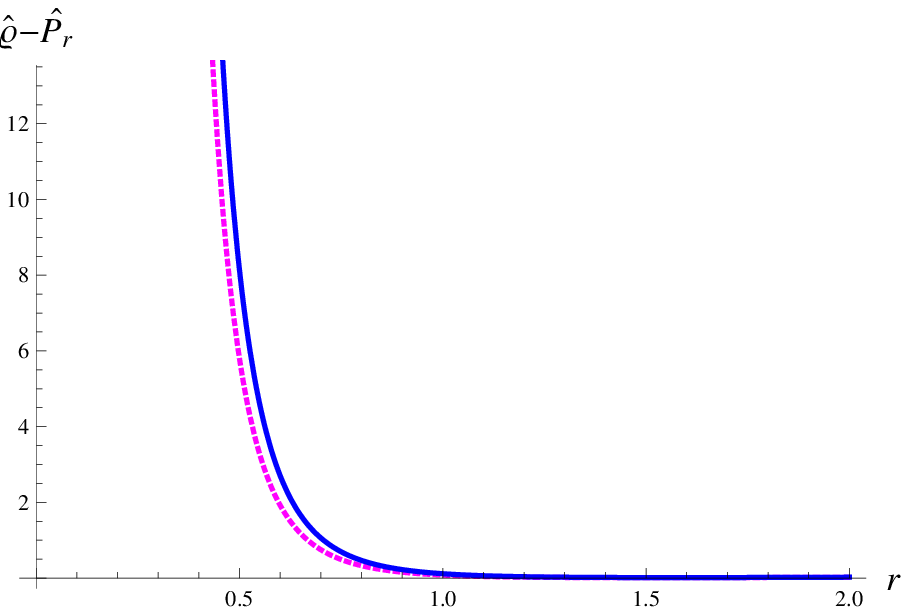,width=0.45\linewidth}
\epsfig{file=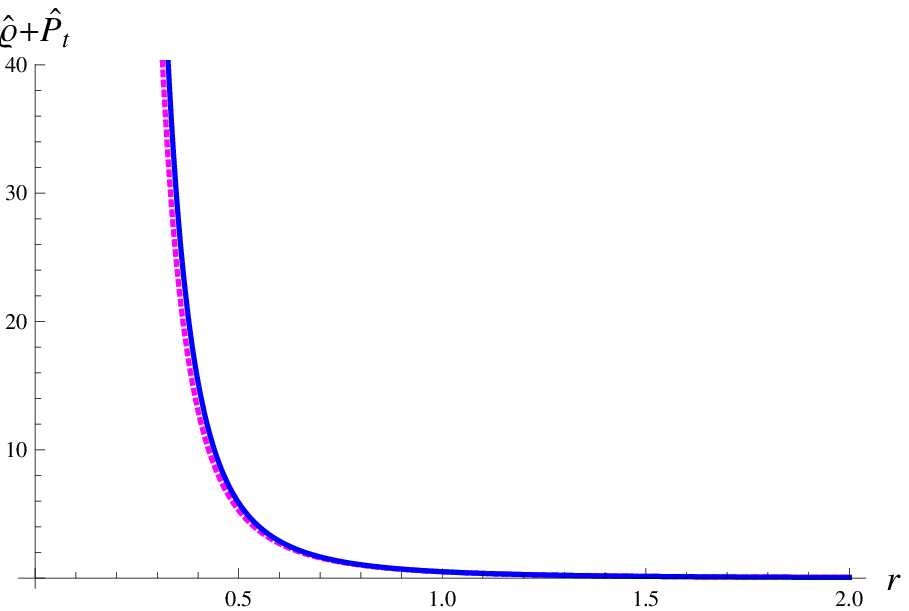,width=0.45\linewidth}\epsfig{file=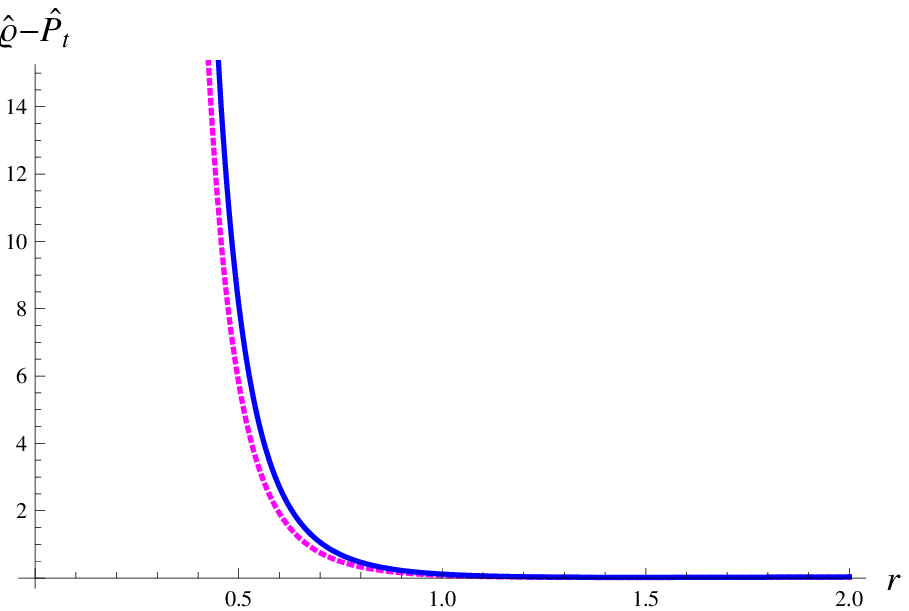,width=0.45\linewidth}
\epsfig{file=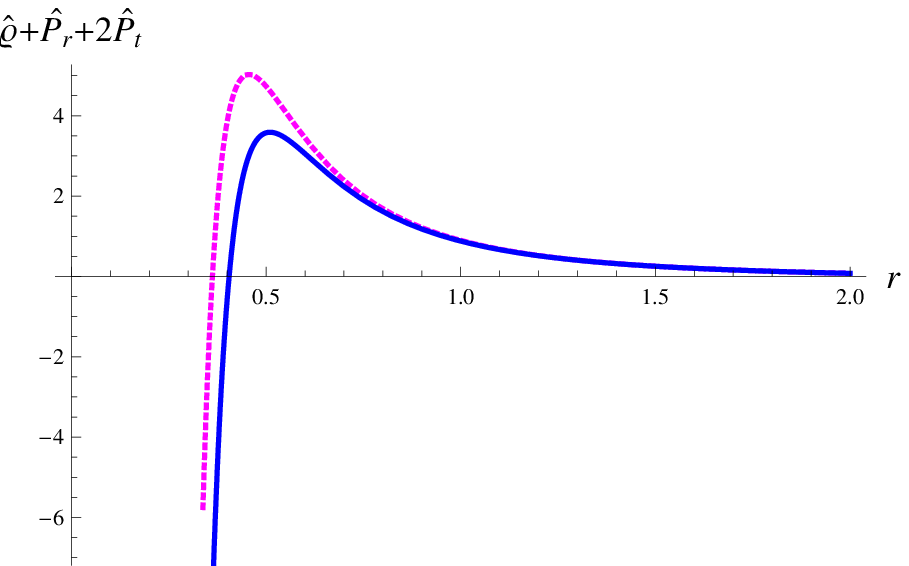,width=0.45\linewidth} \caption{Plots of energy
conditions of the exterior for $\aleph=2.5$ (Magenta) and
$\aleph=3.5$ (Blue).}
\end{figure}

\section{Concluding Remarks}

In this paper, we have constructed anisotropic version of the
gravastar in EMSG by employing gravitational decoupling technique.
The system of field equations is separated into two different
arrays: one belongs to the standard version of $f(\Re, {T}^{2})$
equations, whereas the other involves an additional source. A
gravastar solution describing the ultra-compact structure with
isotropic matter distribution is used to evaluate the first system.
The second set involves four unknowns, i.e., three extra source
components of $\tau_{\sigma\gamma}$ and decoupling function
$\varsigma(r)$. The second system has been closed by employing the
property of gravastars, i.e., $g_{00}=0=g_{11}^{-1}$ at
$\mathfrak{R}=2M$ (Schwarzschild limit). This feature provides the
deformed radial metric component (\ref{104}) for $n\geq2$, which
leads to the deformed version of gravastar given by
Eqs.(\ref{101*}), (\ref{104}) and (\ref{106})-(\ref{108}) in EMSG
gravity.

For the outer geometry, we have assumed two cases. The first case
deals with the standard Schwarzschild geometry whereas the crucial
impact of coupling constant $\zeta$ of the theory is observed
through a relation of the decoupling parameter $\aleph$. This also
ensures the regularity condition which is not found in GR
\cite{12*}, as the smooth matching of the deformed interior with
standard Schwarzschild exterior is not allowed by the regularity
condition. Secondly, the deformed exterior is smoothly matched with
the deformed interior. The density, radial/tagential pressure and
anisotropic factor have acceptable behavior (Figures \textbf{1} and
\textbf{2}). It is mentioned here that the standard Schwarzschild
spacetime has a linear profile while the deformed Schwarzschild
metric follows positive increasing trend (Figure \textbf{3}). The
finite positive behavior of metric potentials (Figure \textbf{4})
has a sharp cusp at the matching point of inner and outer
boundaries. We have also discussed viability of the model through
energy bounds (Figures \textbf{5} and \textbf{6}) and found that all
the energy conditions are satisfied except SEC.

It is worthwhile to mention here that the new version provides an
ultra-compact structure that satisfies the requirement of a viable
celestial body. Finally, we conclude that the gravastar structure is
more compact and stable under the impact of anisotropy in comparison
to the related work in other modified theories of gravity using
isotropic matter configuration \cite{29aa}-\cite{cg77}. Furthermore,
our analysis is found to be consistent with GR \cite{n} and
curvature-matter coupled theory of gravity \cite{n1}.

\end{document}